# Liutex-based Modified Navier-Stokes Equation

Yifei Yu[1], Jianming Liu[2], Yonghua Yan[3], Yiqian Wang[4], Yisheng Gao[5], Chaoqun Liu[1] *

1.  Department of Mathematics, University of Texas at Arlington, Arlington, Texas 76019, USA

2.  School of Mathematics and Statistics, Jiangsu Normal University, Xuzhou 221116, China

3.  Department of Mathematics and Statistics Sciences, Jackson State University, MS 39217, USA

4.  School of Mathematical Science, Soochow University, Suzhou 215006, China

5.  Nanjing University of Aeronautics and Astronautics, Nanjing, China

Abstract:

The Navier-Stokes (NS) partial differential equations, as the governing equation of fluid dynamics, are based on particles with zero volume. In such a condition, conservation law of moment of momentum is automatically satisfied, and thus NS equations only contain conservation of mass, momentum and energy. Because of the difficulty to get an analytical solution of NS equations, scientists develop finite element method (FEM) and finite volume method (FVM) to obtain approximate solutions. However, these methods are dependent on the size of finite volumes, which is not zero. Considering the size of the finite volume, conservation law of moment of momentum is no longer automatically satisfied and it is necessary to develop a new control equation to take effect of momentum into account. In this paper, new relations of reciprocal shear stresses are derived from conservation law of moment of momentum, and so are new constitutive relations and modified NS equations. The Liutex based NS equation may be the universal governing equations for both laminar and turbulent flow which is dominated by vortices and conservation of moment of momentum must be satisfied.

Keywords: conservation of moment of momentum, NS equation, Liutex, inertia

## 1. Introduction

Navier-Stokes equations, as the governing equations of fluid dynamics, only have mass conservation, momentum conservation and energy conservation equations, but not include conservation of moment of momentum. Admittedly, the conservation of moment of momentum is automatically satisfied if the formula is based on infinitesimal particles, which have zero-volume resulting in the moment of inertia to be zero. Getting an analytical solution of Navier-Stokes equation has never been an easy work, because of the complicity and non-linearity of the partial differential equations. As a way to deal with this problem, numerical methods like finite element method (FEM) and finite volume method (FVM) become viable tools. In 1965, Zienkiewicz and Cheung [1] first provided the idea of finite element method. The main idea of FEM is to discretize the computation domain to many finite elements and then apply the governing equations on each element. In the book < Weather prediction by numerical process>, Lewis Fry Richardson[2], together with Richardson, set up the earliest type of computational fluid dynamics(CFD) by finite difference method, extending the application of FEM(FVM) to fluid area. By using CFD method, approximate solution can be obtained for almost all practical problems. However, the main difference between FEM or FVM and the theoretical governing equations is that FEM is no longer based on particles, but elements with finite volume, which makes conservation of moment of momentum not satisfied. This becomes an extreme problem in the analysis of turbulence, since there are a lot of rotations in turbulence, meaning the moment of momentum plays a role that cannot be ignored. Launder and Spalding[3] created $k$-$\varepsilon$ model to solve this problem, and many other models are established in the past decades, including S-A model[4], $k$-$\omega$ model[5], SST model[6] and etc. However, these modules only can solve a particular type of cases, none of them is applied uniformly. So, in this paper, a modified NS equation with conservation of moment of momentum will be established to solve the problem with conservation of moment of momentum. Since the new modified NS equation is based on conservation of moment of momentum, it has great possibility to be a uniform governing equation.

*Corresponding author: Chaoqun Liu, cliu@uta.edu

Conservation of moment of momentum is a relation linking torque and angular acceleration. Therefore, to use this conservation law, a physical quantity is needed to estimate the fluid rotation. In 2018, Liu et al[7] defined mathematically a physical quantity, which is called Liutex, to represent fluid rotation or vortex. Liutex is a vector[8], whose magnitude is twice of the angular speed and whose direction is the swirling axis. The reason why not using the well-known vorticity, Q criterion[9] or $\lambda_{ci}$ method[10] is that these methods are more or less contaminated by shear or stretching or both, and Liutex extracts the rigid rotation part from fluid motion.

This paper is organized as follows. In Section 2, new relations of reciprocal shear stresses are derived by applying conservation law of moment of momentum, followed by new constitutive relations in Section 3. The modified NS equations are given in Section 4 and there is a conclusion in Section 5.

## 2. Reciprocal shear stress relations

In this section, conservation law of moment of momentum is applied to derive a new control equation for fluid dynamics.

[a]Email: cliu@uta.edu

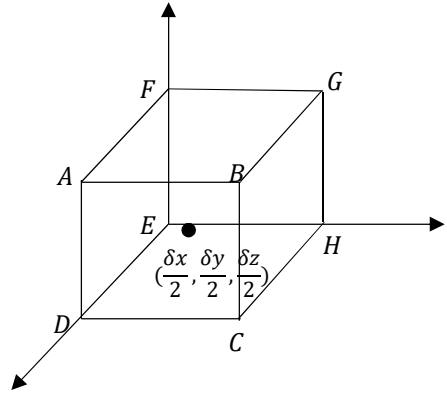

Figure 1 Control volume

For simplicity, it is assumed the control volume is a cubic and the coordinate is established as shown in the Figure 1. The center of the cubic is marked as $\left(\dfrac{\delta x}{2}, \dfrac{\delta y}{2}, \dfrac{\delta z}{2}\right)$ and $\delta x, \delta y, \delta z$ are the length, width and height of the control cuboid, respectively.

The first order Taylor expansion of shear stress is written as:

$$\tau_{xy}(x,y,z) = \tau_{xy}(\frac{\delta x}{2}, \frac{\delta y}{2}, \frac{\delta z}{2}) + A_1(x - \frac{\delta x}{2}) + A_2(y - \frac{\delta y}{2}) + A_3(z - \frac{\delta z}{2}) \qquad (1)$$

Where

$$A_1 = \frac{\partial \tau_{xy}\left(\dfrac{\delta x}{2}, \dfrac{\delta y}{2}, \dfrac{\delta z}{2}\right)}{\partial x}, A_2 = \frac{\partial \tau_{xy}\left(\dfrac{\delta x}{2}, \dfrac{\delta y}{2}, \dfrac{\delta z}{2}\right)}{\partial y}, A_3 = \frac{\partial \tau_{xy}\left(\dfrac{\delta x}{2}, \dfrac{\delta y}{2}, \dfrac{\delta z}{2}\right)}{\partial z} \qquad (2)$$

Similarly, the Taylor expressions of other shear stresses are:

$$\tau_{yx}(x,y,z) = \tau_{yx}(\frac{\delta x}{2}, \frac{\delta y}{2}, \frac{\delta z}{2}) + B_1(x - \frac{\delta x}{2}) + B_2(y - \frac{\delta y}{2}) + B_3(z - \frac{\delta z}{2}) \qquad (3)$$

$$\tau_{xz}(x,y,z) = \tau_{xz}(\frac{\delta x}{2}, \frac{\delta y}{2}, \frac{\delta z}{2}) + C_1(x - \frac{\delta x}{2}) + C_2(y - \frac{\delta y}{2}) + C_3(z - \frac{\delta z}{2}) \qquad (4)$$

$$\tau_{zx}(x,y,z) = \tau_{zx}(\frac{\delta x}{2}, \frac{\delta y}{2}, \frac{\delta z}{2}) + D_1(x - \frac{\delta x}{2}) + D_2(y - \frac{\delta y}{2}) + D_3(z - \frac{\delta z}{2}) \qquad (5)$$

$$\tau_{yz}(x,y,z) = \tau_{yz}(\frac{\delta x}{2}, \frac{\delta y}{2}, \frac{\delta z}{2}) + E_1(x - \frac{\delta x}{2}) + E_2(y - \frac{\delta y}{2}) + E_3(z - \frac{\delta z}{2}) \qquad (6)$$

$$\tau_{zy}(x,y,z) = \tau_{zy}(\frac{\delta x}{2}, \frac{\delta y}{2}, \frac{\delta z}{2}) + F_1(x - \frac{\delta x}{2}) + F_2(y - \frac{\delta y}{2}) + F_3(z - \frac{\delta z}{2}) \qquad (7)$$

$$\tau_{xx}(x,y,z) = \tau_{xx}(\frac{\delta x}{2}, \frac{\delta y}{2}, \frac{\delta z}{2}) + H_1(x - \frac{\delta x}{2}) + H_2(y - \frac{\delta y}{2}) + H_3(z - \frac{\delta z}{2}) \tag{8}$$

$$\tau_{yy}(x,y,z) = \tau_{yy}(\frac{\delta x}{2}, \frac{\delta y}{2}, \frac{\delta z}{2}) + J_1(x - \frac{\delta x}{2}) + J_2(y - \frac{\delta y}{2}) + J_3(z - \frac{\delta z}{2}) \tag{9}$$

$$\tau_{zz}(x,y,z) = \tau_{zz}(\frac{\delta x}{2}, \frac{\delta y}{2}, \frac{\delta z}{2}) + K_1(x - \frac{\delta x}{2}) + K_2(y - \frac{\delta y}{2}) + K_3(z - \frac{\delta z}{2}) \tag{10}$$

Where

$$B_1 = \frac{\partial \tau_{yx}\left(\frac{\delta x}{2}, \frac{\delta y}{2}, \frac{\delta z}{2}\right)}{\partial x}, B_2 = \frac{\partial \tau_{yx}\left(\frac{\delta x}{2}, \frac{\delta y}{2}, \frac{\delta z}{2}\right)}{\partial y}, B_3 = \frac{\partial \tau_{yx}\left(\frac{\delta x}{2}, \frac{\delta y}{2}, \frac{\delta z}{2}\right)}{\partial z} \tag{11}$$

$$C_1 = \frac{\partial \tau_{xz}\left(\frac{\delta x}{2}, \frac{\delta y}{2}, \frac{\delta z}{2}\right)}{\partial x}, C_2 = \frac{\partial \tau_{xz}\left(\frac{\delta x}{2}, \frac{\delta y}{2}, \frac{\delta z}{2}\right)}{\partial y}, C_3 = \frac{\partial \tau_{xz}\left(\frac{\delta x}{2}, \frac{\delta y}{2}, \frac{\delta z}{2}\right)}{\partial z} \tag{12}$$

$$D_1 = \frac{\partial \tau_{zx}\left(\frac{\delta x}{2}, \frac{\delta y}{2}, \frac{\delta z}{2}\right)}{\partial x}, D_2 = \frac{\partial \tau_{zx}\left(\frac{\delta x}{2}, \frac{\delta y}{2}, \frac{\delta z}{2}\right)}{\partial y}, D_3 = \frac{\partial \tau_{zx}\left(\frac{\delta x}{2}, \frac{\delta y}{2}, \frac{\delta z}{2}\right)}{\partial z} \tag{13}$$

$$E_1 = \frac{\partial \tau_{yz}\left(\frac{\delta x}{2}, \frac{\delta y}{2}, \frac{\delta z}{2}\right)}{\partial x}, E_2 = \frac{\partial \tau_{yz}\left(\frac{\delta x}{2}, \frac{\delta y}{2}, \frac{\delta z}{2}\right)}{\partial y}, E_3 = \frac{\partial \tau_{yz}\left(\frac{\delta x}{2}, \frac{\delta y}{2}, \frac{\delta z}{2}\right)}{\partial z} \tag{14}$$

$$F_1 = \frac{\partial \tau_{zy}\left(\frac{\delta x}{2}, \frac{\delta y}{2}, \frac{\delta z}{2}\right)}{\partial x}, F_2 = \frac{\partial \tau_{zy}\left(\frac{\delta x}{2}, \frac{\delta y}{2}, \frac{\delta z}{2}\right)}{\partial y}, F_3 = \frac{\partial \tau_{zy}\left(\frac{\delta x}{2}, \frac{\delta y}{2}, \frac{\delta z}{2}\right)}{\partial z} \tag{15}$$

$$H_1 = \frac{\partial \tau_{xx}\left(\frac{\delta x}{2}, \frac{\delta y}{2}, \frac{\delta z}{2}\right)}{\partial x}, H_2 = \frac{\partial \tau_{xx}\left(\frac{\delta x}{2}, \frac{\delta y}{2}, \frac{\delta z}{2}\right)}{\partial y}, H_3 = \frac{\partial \tau_{xx}\left(\frac{\delta x}{2}, \frac{\delta y}{2}, \frac{\delta z}{2}\right)}{\partial z} \tag{16}$$

$$J_1 = \frac{\partial \tau_{yy}\left(\frac{\delta x}{2}, \frac{\delta y}{2}, \frac{\delta z}{2}\right)}{\partial x}, J_2 = \frac{\partial \tau_{yy}\left(\frac{\delta x}{2}, \frac{\delta y}{2}, \frac{\delta z}{2}\right)}{\partial y}, J_3 = \frac{\partial \tau_{yy}\left(\frac{\delta x}{2}, \frac{\delta y}{2}, \frac{\delta z}{2}\right)}{\partial z} \tag{17}$$

$$K_1 = \frac{\partial \tau_{zz}\left(\frac{\delta x}{2}, \frac{\delta y}{2}, \frac{\delta z}{2}\right)}{\partial x}, K_2 = \frac{\partial \tau_{zz}\left(\frac{\delta x}{2}, \frac{\delta y}{2}, \frac{\delta z}{2}\right)}{\partial y}, K_3 = \frac{\partial \tau_{zz}\left(\frac{\delta x}{2}, \frac{\delta y}{2}, \frac{\delta z}{2}\right)}{\partial z} \tag{18}$$

Then, moment of momentum conservation law is applied to z-direction (Figure 2). Firstly, expression of shear stresses on each surface are figured out.

[a]Email: cliu@uta.edu

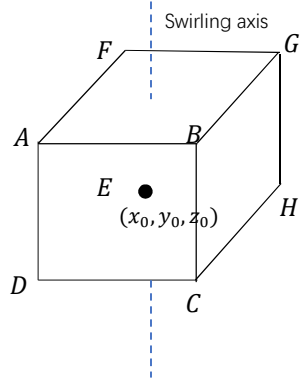

Figure 2 Swirling axis in z-direction

The positive directions of the stresses are shown in Figure 3.

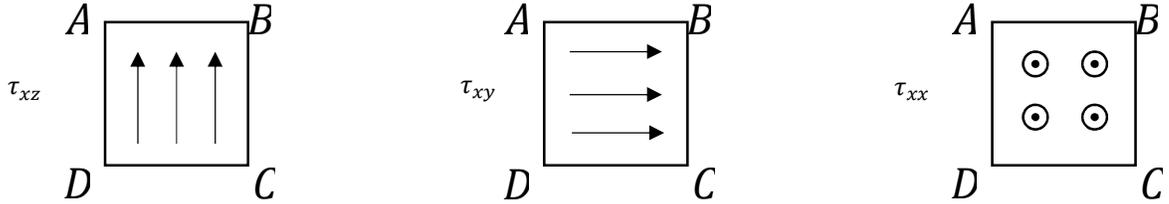

Figure 3. Shear stresses on one surface

The direction of $\tau_{xz}$ is parallel to z direction which leads to its force arm to be zero and the torque to be zero. The torque caused by $\tau_{xy}$ can be expressed as the following integral

$$
\begin{aligned}
& T_{ABCDxy} \\
& = \iint_{S_{ABCD}} \tau_{xy}(\delta x, y, z)\frac{\delta x}{2}dydz \\
& = \iint_{S_{ABCD}} \left[\tau_{xy}(\frac{\delta x}{2},\frac{\delta y}{2},\frac{\delta z}{2}) + A_1(\delta x - \frac{\delta x}{2}) + A_2(y - \frac{\delta y}{2}) + A_3(z - \frac{\delta z}{2})\right]\frac{\delta x}{2}dydz \\
& = \int_0^{\delta z}\int_0^{\delta y} \tau_{xy}(\frac{\delta x}{2},\frac{\delta y}{2},\frac{\delta z}{2})\frac{\delta x}{2}dydz + \int_0^{\delta z}\int_0^{\delta y} A_1\frac{\delta x}{2}\frac{\delta x}{2}dydz + \int_0^{\delta z}\int_0^{\delta y} A_2(y - \frac{\delta y}{2})\frac{\delta x}{2}dydz + \int_0^{\delta z}\int_0^{\delta y} A_3(z - \frac{\delta z}{2})\frac{\delta x}{2}dydz
\end{aligned}
$$

(19)

Since

[a]Email: cliu@uta.edu

$$\int\limits_{0}^{\delta z}\int\limits_{0}^{\delta y} A_2(y-\frac{\delta y}{2})\frac{\delta x}{2}dydz = \int\limits_{0}^{\delta z}\int\limits_{0}^{\delta y} A_3(z-\frac{\delta z}{2})\frac{\delta x}{2}dydz = 0 \tag{20}$$

This expression can be simplified as

$$
\begin{aligned}
T_{ABCDxy} \\
&= \int\limits_{0}^{\delta z}\int\limits_{0}^{\delta y} \tau_{xy}(\frac{\delta x}{2},\frac{\delta y}{2},\frac{\delta z}{2})\frac{\delta x}{2}dydz + \int\limits_{0}^{\delta z}\int\limits_{0}^{\delta y} A_1\frac{\delta x}{2}\frac{\delta x}{2}dydz \\
&= \frac{1}{2}\tau_{xy}(\frac{\delta x}{2},\frac{\delta y}{2},\frac{\delta z}{2})\delta x\delta y\delta z + \frac{1}{4}A_1(\delta x)^2\delta y\delta z
\end{aligned}
\tag{21}
$$

The torque caused by $\tau_{xx}$ can be evaluated by the following integral.

$$
\begin{aligned}
T_{ABCDxx} \\
&= \iint\limits_{S_{ABCD}} \tau_{xx}(\delta x,y,z)\left(\frac{\delta y}{2}-y\right)dydz \\
&= \iint\limits_{S_{ABCD}} \left[\tau_{xx}(\frac{\delta x}{2},\frac{\delta y}{2},\frac{\delta z}{2})+H_1(\delta x-\frac{\delta x}{2})+H_2(y-\frac{\delta y}{2})+H_3(z-\frac{\delta z}{2})\right]\left(\frac{\delta y}{2}-y\right)dydz \\
&= \int\limits_{0}^{\delta z}\int\limits_{0}^{\delta y} \tau_{xx}(\frac{\delta x}{2},\frac{\delta y}{2},\frac{\delta z}{2})\left(\frac{\delta y}{2}-y\right)dydz + \int\limits_{0}^{\delta z}\int\limits_{0}^{\delta y} H_1\frac{\delta x}{2}\frac{\delta x}{2}\left(\frac{\delta y}{2}-y\right)dydz \\
&\quad + \int\limits_{0}^{\delta z}\int\limits_{0}^{\delta y} H_2(y-\frac{\delta y}{2})\left(\frac{\delta y}{2}-y\right)dydz + \int\limits_{0}^{\delta z}\int\limits_{0}^{\delta y} H_3(z-\frac{\delta z}{2})\left(\frac{\delta y}{2}-y\right)dydz
\end{aligned}
\tag{22}
$$

Easy to know,

$$\int\limits_{0}^{\delta z}\int\limits_{0}^{\delta y}\tau_{xx}\frac{\delta x}{2},\frac{\delta y}{2},\frac{\delta z}{2}\left(\frac{\delta y}{2}-y\right)dydz = \int\limits_{0}^{\delta z}\int\limits_{0}^{\delta y} H_1\frac{\delta x}{2}\frac{\delta x}{2}\left(\frac{\delta y}{2}-y\right)dydz = \int\limits_{0}^{\delta z}\int\limits_{0}^{\delta y} H_3(z-\frac{\delta z}{2})\left(\frac{\delta y}{2}-y\right)dydz = 0 \tag{23}$$

Therefore,

$$T_{ABCDxx} = \int\limits_{0}^{\delta z}\int\limits_{0}^{\delta y} H_2(y-\frac{\delta y}{2})\left(\frac{\delta y}{2}-y\right)dydz = -\frac{1}{6}H_2(\delta y)^3\delta z \tag{24}$$

a)Email: cliu@uta.edu

Next, the torques caused by stresses on other surfaces are estimated. The process is similar to what we did about surface ABCD, so here, the trivial steps are skipped and the expressions are given directly.

$$T_{FGHExy} = \frac{1}{2}\tau_{xy}(\frac{\delta x}{2}, \frac{\delta y}{2}, \frac{\delta z}{2})\delta x \delta y \delta z - \frac{1}{4}A_1(\delta x)^2 \delta y \delta z \tag{25}$$

$$T_{FGHExx} = \frac{1}{6}H_2(\delta y)^3 \delta z \tag{26}$$

$$T_{ADEFyx} = -\frac{1}{2}\tau_{yx}(\frac{\delta x}{2}, \frac{\delta y}{2}, \frac{\delta z}{2})\delta x \delta y \delta z + \frac{1}{4}B_2 \delta x (\delta y)^2 \delta z \tag{27}$$

$$T_{ADEFyy} = -\frac{1}{6}J_1(\delta x)^3 \delta z \tag{28}$$

$$T_{BGHCyx} = -\frac{1}{2}\tau_{yx}(\frac{\delta x}{2}, \frac{\delta y}{2}, \frac{\delta z}{2})\delta x \delta y \delta z - \frac{1}{4}B_2 \delta x (\delta y)^2 \delta z \tag{29}$$

$$T_{BGHCyy} = \frac{1}{6}J_1(\delta x)^3 \delta z \tag{30}$$

$$T_{AFGBzx} = -\frac{1}{6}D_2 \delta x (\delta y)^3 \tag{31}$$

$$T_{AFGBzy} = \frac{1}{6}F_1(\delta x)^3 \delta y \tag{32}$$

$$T_{CDEHzx} = \frac{1}{6}D_2 \delta x (\delta y)^3 \tag{33}$$

$$T_{CDEHzy} = -\frac{1}{6}F_1(\delta x)^3 \delta y \tag{34}$$

Therefore, the total torque relative to z-axis is

[a]Email: cliu@uta.edu

$$T = T_{ABCDxy} + T_{ABCDxx} + T_{FGHExy} + T_{FGHExx} + T_{ADEFyx} + T_{ADEFyy}$$

$$+ T_{BGHCyx} + T_{BGHCyy} + T_{AFGBzx} + T_{AFGBzy} + T_{CDEHzx} + T_{CDEHzy}$$

$$= \frac{1}{2}\tau_{xy}(\frac{\delta x}{2}, \frac{\delta y}{2}, \frac{\delta z}{2})\delta x \delta y \delta z + \frac{1}{4}A_1(\delta x)^2 \delta y \delta z - \frac{1}{6}H_2(\delta y)^3 \delta z$$

$$+ \frac{1}{2}\tau_{xy}(\frac{\delta x}{2}, \frac{\delta y}{2}, \frac{\delta z}{2})\delta x \delta y \delta z - \frac{1}{4}A_1(\delta x)^2 \delta y \delta z + \frac{1}{6}H_2(\delta y)^3 \delta z$$

$$- \frac{1}{2}\tau_{yx}(\frac{\delta x}{2}, \frac{\delta y}{2}, \frac{\delta z}{2})\delta x \delta y \delta z + \frac{1}{4}B_2 \delta x(\delta y)^2 \delta z - \frac{1}{6}J_1(\delta x)^3 \delta z$$

$$- \frac{1}{2}\tau_{yx}(\frac{\delta x}{2}, \frac{\delta y}{2}, \frac{\delta z}{2})\delta x \delta y \delta z - \frac{1}{4}B_2 \delta x(\delta y)^2 \delta z + \frac{1}{6}J_1(\delta x)^3 \delta z$$

$$- \frac{1}{6}D_2 \delta x(\delta y)^3 + \frac{1}{6}F_1(\delta x)^3 \delta y + \frac{1}{6}D_2 \delta x(\delta x)^3 - \frac{1}{6}F_1(\delta x)^3 \delta y$$

$$= \tau_{xy}(\frac{\delta x}{2}, \frac{\delta y}{2}, \frac{\delta z}{2})\delta x \delta y \delta z - \tau_{yx}(\frac{\delta x}{2}, \frac{\delta y}{2}, \frac{\delta z}{2})\delta x \delta y \delta z$$

$$(35)$$

Let $I_z$ be the moment of inertia relative to z-axis, and $\frac{R_z}{2}$ represents the angular speed in the z-direction.

Then, the moment of momentum conservation equation can be written as

$$I_z \frac{\partial R_z}{2\partial t} = \left[\tau_{xy}(\frac{\delta x}{2}, \frac{\delta y}{2}, \frac{\delta z}{2}) - \tau_{yx}(\frac{\delta x}{2}, \frac{\delta y}{2}, \frac{\delta z}{2})\right]\delta x \delta y \delta z \qquad (36)$$

Denote $\gamma = \frac{I_z}{4\delta x \delta y \delta z}$, then the above relation can be expressed as

$$\tau_{xy} = \tau_{yx} + 2\gamma \frac{\partial R_z}{\partial t} \quad , \qquad (37)$$

where $R_z$ is the z-component of the Liutex vector.

We can get relations between the other two pairs of shear stresses similarly, and the three relations between pairs of shear stresses are:

$$\tau_{yz} = \tau_{zy} + 2\alpha \frac{\partial R_x}{\partial t} \qquad (38)$$

$$\tau_{zx} = \tau_{xz} + 2\beta \frac{\partial R_y}{\partial t} \qquad (39)$$

$$\tau_{xy} = \tau_{yx} + 2\gamma \frac{\partial R_z}{\partial t} \qquad (40)$$

[a]Email: cliu@uta.edu

where $\alpha = \dfrac{I_x}{4\delta x \delta y \delta z}$, $\beta = \dfrac{I_y}{4\delta x \delta y \delta z}$, and $\gamma = \dfrac{I_z}{4\delta x \delta y \delta z}$. $I_x$, $I_y$ and $I_z$ are moment of

inertia relative to x, y and z-axis respectively.

From Eqs. (38), (39) and (40), it can be seen that reciprocal shear stresses are not equal to each other. The reciprocal theorem of shear stress clearly states reciprocal shear stresses are equal to each other. Does it mean the reciprocal theorem of shear stress is incorrect? Definitely not and this theorem is correct, but the key issue is reciprocal theorem is based on a particle with infinitesimal volume. However, the above derived relations depend on the size of a finite volume. If the volume approximates to 0, then $\alpha = \beta = \gamma = 0$ and these three relations become exactly the same as reciprocal theorem of shear stress.

3. New constitutive relations

Recall the classical constitutive relations[11] which are:

$$\tau_{yz} = \tau_{zy} = \mu\left(\frac{\partial v}{\partial z} + \frac{\partial w}{\partial y}\right) \tag{41}$$

$$\tau_{zx} = \tau_{xz} = \mu\left(\frac{\partial u}{\partial z} + \frac{\partial w}{\partial x}\right) \tag{42}$$

$$\tau_{xy} = \tau_{yx} = \mu\left(\frac{\partial u}{\partial y} + \frac{\partial v}{\partial x}\right) \tag{43}$$

where $\mu$ is viscosity.

These relations are widely used in fluid dynamics, and they are considered very close to the right answer, although they assume reciprocal shear stresses are equal. So, the new constitutive relations come from a little modification of the classical ones to make them satisfy relations (38), (39) and (40). In this paper, we distribute their difference evenly to each one. The new constitutive relations are:

$$\tau_{yz} = \mu\left(\frac{\partial v}{\partial z} + \frac{\partial w}{\partial y}\right) + \alpha \frac{\partial R_x}{\partial t} \tag{44}$$

$$\tau_{zy} = \mu\left(\frac{\partial v}{\partial z} + \frac{\partial w}{\partial y}\right) - \alpha \frac{\partial R_x}{\partial t} \tag{45}$$

$$\tau_{zx} = \mu\left(\frac{\partial u}{\partial z} + \frac{\partial w}{\partial x}\right) + \beta\frac{\partial R_y}{\partial t} \tag{46}$$

$$\tau_{xz} = \mu\left(\frac{\partial u}{\partial z} + \frac{\partial w}{\partial x}\right) - \beta\frac{\partial R_y}{\partial t} \tag{47}$$

$$\tau_{yy} = \mu\left(\frac{\partial u}{\partial y} + \frac{\partial v}{\partial x}\right) + \gamma\frac{\partial R_z}{\partial t} \tag{48}$$

$$\tau_{yx} = \mu\left(\frac{\partial u}{\partial y} + \frac{\partial v}{\partial x}\right) - \gamma\frac{\partial R_z}{\partial t} \tag{49}$$

By adding one term and subtracting the same term, the 6 new constitutive relations satisfy the requirements (38), (39) and (40). The new constitutive relations are more accurate than the classical ones when doing research on finite volumes with nonzero size, because the influence of volume size is taken into consideration.

4.  Modified NS equation

With new constitutive relations in hand, modifications can be manipulated on the traditional Navier-Stokes equation. We assume the viscosity is a constant.

$$\begin{aligned}\frac{\partial \tau_{xx}}{\partial x} &= \frac{\partial\left[-\dfrac{2}{3}\mu\left(\dfrac{\partial u}{\partial x} + \dfrac{\partial v}{\partial y} + \dfrac{\partial w}{\partial z}\right) + 2\mu\dfrac{\partial u}{\partial x} - p\right]}{\partial x}\\ &= -\frac{2}{3}\mu\left(\frac{\partial^2 u}{\partial x^2} + \frac{\partial^2 v}{\partial x\partial y} + \frac{\partial^2 w}{\partial x\partial z}\right) + 2\mu\frac{\partial^2 u}{\partial x^2} - \frac{\partial P}{\partial x}\end{aligned} \tag{50}$$

$$\frac{\partial \tau_{yx}}{\partial y} = \frac{\partial}{\partial y}\left[\mu\left(\frac{\partial u}{\partial y} + \frac{\partial v}{\partial x}\right) - \gamma\frac{\partial R_z}{\partial t}\right] = \mu\left(\frac{\partial^2 u}{\partial y^2} + \frac{\partial^2 v}{\partial x\partial y}\right) - \gamma\frac{\partial^2 R_z}{\partial y\partial t} \tag{51}$$

$$\frac{\partial \tau_{zx}}{\partial z} = \frac{\partial}{\partial z}\left[\mu\left(\frac{\partial u}{\partial z} + \frac{\partial w}{\partial x}\right) + \beta\frac{\partial R_y}{\partial t}\right] = \mu\left(\frac{\partial^2 u}{\partial z^2} + \frac{\partial^2 w}{\partial x\partial z}\right) + \beta\frac{\partial^2 R_y}{\partial z\partial t} \tag{52}$$

Plug (50), (51) and (52) into the following NS equation in x direction[12]

$$\frac{\partial(\rho u)}{\partial t} + \frac{\partial(\rho uu)}{\partial x} + \frac{\partial(\rho vu)}{\partial y} + \frac{\partial(\rho wu)}{\partial z} = \rho f_x + \frac{\partial \tau_{xx}}{\partial x} + \frac{\partial \tau_{yx}}{\partial y} + \frac{\partial \tau_{zx}}{\partial z} \tag{53}$$

[a]Email: cliu@uta.edu

Then, it becomes

$$\frac{\partial(\rho u)}{\partial t} + \frac{\partial(\rho uu)}{\partial x} + \frac{\partial(\rho vu)}{\partial y} + \frac{\partial(\rho wu)}{\partial z} = \rho f_x - \frac{2}{3}\mu\left(\frac{\partial^2 u}{\partial x^2} + \frac{\partial^2 v}{\partial x\partial y} + \frac{\partial^2 w}{\partial x\partial z}\right) + 2\mu\frac{\partial^2 u}{\partial x^2}$$
$$-\frac{\partial P}{\partial x} + \mu\left(\frac{\partial^2 u}{\partial y^2} + \frac{\partial^2 v}{\partial x\partial y}\right) - \gamma\frac{\partial^2 R_z}{\partial y\partial t} + \mu\left(\frac{\partial^2 u}{\partial z^2} + \frac{\partial^2 w}{\partial x\partial z}\right) + \beta\frac{\partial^2 R_y}{\partial z\partial t} \qquad (54)$$

$$\frac{\partial(\rho u)}{\partial t} + \frac{\partial(\rho uu)}{\partial x} + \frac{\partial(\rho vu)}{\partial y} + \frac{\partial(\rho wu)}{\partial z} = \rho f_x - \frac{2}{3}\mu\frac{\partial}{\partial x}\left(\frac{\partial u}{\partial x} + \frac{\partial v}{\partial y} + \frac{\partial w}{\partial z}\right)$$
$$+\mu\left(\frac{\partial^2 u}{\partial x^2} + \frac{\partial^2 u}{\partial y^2} + \frac{\partial^2 u}{\partial z^2}\right) + \mu\frac{\partial}{\partial x}\left(\frac{\partial u}{\partial x} + \frac{\partial v}{\partial y} + \frac{\partial w}{\partial z}\right) - \frac{\partial P}{\partial x} - \gamma\frac{\partial^2 R_z}{\partial y\partial t} + \beta\frac{\partial^2 R_y}{\partial z\partial t} \qquad (55)$$

Assume the fluid is incompressible, i.e.

$$\frac{\partial u}{\partial x} + \frac{\partial v}{\partial y} + \frac{\partial w}{\partial z} = 0 \qquad (56)$$

Then, (55) can be simplified as

$$\frac{\partial(\rho u)}{\partial t} + \frac{\partial(\rho uu)}{\partial x} + \frac{\partial(\rho vu)}{\partial y} + \frac{\partial(\rho wu)}{\partial z} = \rho f_x + \mu\left(\frac{\partial^2 u}{\partial x^2} + \frac{\partial^2 u}{\partial y^2} + \frac{\partial^2 u}{\partial z^2}\right) - \frac{\partial P}{\partial x} - \gamma\frac{\partial^2 R_z}{\partial y\partial t} + \beta\frac{\partial^2 R_y}{\partial z\partial t} \qquad (57)$$

Similarly, equations in y- and z- directions can be derived.

$$\frac{\partial(\rho v)}{\partial t} + \frac{\partial(\rho uv)}{\partial x} + \frac{\partial(\rho vv)}{\partial y} + \frac{\partial(\rho wv)}{\partial z} = \rho f_y + \mu\left(\frac{\partial^2 v}{\partial x^2} + \frac{\partial^2 v}{\partial y^2} + \frac{\partial^2 v}{\partial z^2}\right) - \frac{\partial P}{\partial y} + \gamma\frac{\partial^2 R_z}{\partial x\partial t} - \alpha\frac{\partial^2 R_x}{\partial z\partial t} \qquad (58)$$

$$\frac{\partial(\rho w)}{\partial t} + \frac{\partial(\rho uw)}{\partial x} + \frac{\partial(\rho vw)}{\partial y} + \frac{\partial(\rho ww)}{\partial z} = \rho f_z + \mu\left(\frac{\partial^2 w}{\partial x^2} + \frac{\partial^2 w}{\partial y^2} + \frac{\partial^2 w}{\partial z^2}\right) - \frac{\partial P}{\partial z} + \alpha\frac{\partial^2 R_x}{\partial y\partial t} - \beta\frac{\partial^2 R_y}{\partial x\partial t} \qquad (59)$$

By now, the modified Navier-Stokes equations have been derived, which satisfies the conservation law of moment of momentum in a finite volume.

## 5. Conclusion

Conservation law of moment of momentum is automatically satisfied if the research is based on an infinitesimal particle, which is assumed to be zero in volume and thus with zero moment of inertia. However, in all engineering computations, methods like FEM or FVM methods, which use finite volume element rather than zero-volume particles to analyze science and engineering problems. Therefore, the new modified NS equations have a great potential to improve the prediction ability of such methods, including improve the accuracy, calculation cost and etc. More important, the new modified Navier-Stokes equation may be the universal governing equation for both laminar flow and turbulent flow which is dominated by vortices or fluid rotation and the conservation of moment of momentum must be considered.

[a]Email: cliu@uta.edu

[a]Email: cliu@uta.edu